\documentclass[fleqn,12pt,twoside]{article}
\usepackage{espcrc1}

\readRCS
$Id: espcrc1.tex,v 1.2 2004/02/24 11:22:11 spepping Exp $
\ifx\pdftexversion\undefined
  \usepackage[dvips]{graphicx}
\else
  \usepackage[pdftex]{graphicx}
\fi
\newcommand{\AmS}{{\protect\the\textfont2
  A\kern-.1667em\lower.5ex\hbox{M}\kern-.125emS}}

\title{Two and Three Particle Flavor Dependent Correlations}

\author{N. N. Ajitanand\address[sunysb]{Dept. of Chemistry, 
        SUNY Stony Brook, Stony Brook, NY 11794, USA}, 
        for the PHENIX Collaboration }
       
\begin{document}

\maketitle

\begin{abstract}
%
The PHENIX collaboration has developed novel methodologies for reliable extraction  
of jet functions from two and three particle azimuthal correlation functions  
measured at mid-rapidity in Au+Au collisions at $\sqrt s=200$ GeV. The extracted jet 
shape and the yield of jet-associated partner hadrons (per trigger hadron) are found to 
vary with particle type and collision centrality, indicating a significant effect 
of the nuclear collision medium on the (di)jet fragmentation process.
\end{abstract}

\section{Introduction}

It is well known that the high energy density achieved in semi-central Au + Au 
collisions at RHIC, far exceeds the lattice QCD estimate for creating a de-confined phase of 
quarks and gluons (QGP)\cite{Phenix_et}. It's rapid thermalization gives rise to large pressure 
gradients evidenced by the sizable azimuthal anisotropy ($v_2$) observed for particle emissions from 
the collision zone \cite{RLacey_QM05}. The magnitude of this anisotropy is compatible with the predictions of 
the hydrodynamic model which in turn implies the creation of a strongly interacting medium 
and essentially full local thermal equilibrium. In addition to the soft processes giving 
rise to the formation of the medium, there are relatively rare hard parton-parton collisions. 
The scattered partons may propagate through the medium radiating gluons and 
interacting with the medium till they finally fragment into jet-like 
clusters. Since such interactions can modify jet fragmentation, jets provide a 
powerful probe of the medium, provided one can reliably extract the jet signal from 
the relatively large background which exist in RHIC collisions. Possible medium 
associated modifications of the jet properties are a shock wave induced conical flow or 
``sonic boom'' \cite{Casalderrey_04} and ``bending" induced by interactions between 
the propagating partons and the flowing medium \cite{Armesto_04}. 

\section{Methodology and results: }

{\bf \em Correlation Functions:} To study jet properties we use two- and 
three-particle azimuthal correlation functions.
These azimuthal correlation functions were built by pairing a 
leading hadron in a specified transverse momentum range $p_T(\mathrm{trig})$, with an 
associated hadron also in a specified range $p_T( \mathrm{assoc})$. For two-particle 
correlations, the correlation function $C\left( {\Delta \phi } \right)$ is given by:
\[
C\left( {\Delta \phi } \right)=\frac{N_{real} \left( {\Delta \phi } 
\right)}{N_{mix} \left( {\Delta \phi } \right)},
\]
where ${\Delta \phi }$ is the difference of the azimuthal angles of the pair. The 
real distribution ($N_{real}\left( {\Delta \phi }\right)$) is built from pair members 
belonging to the same event and the mixed distribution ($N_{mix}\left( {\Delta \phi }\right)$) 
is made of pair members belonging to different events. Thus the correlation function is free 
of geometric acceptance effects and carries only the combined correlations from flow and jets. 
Decomposition of these correlations into their jet and flow contributions, constitute an 
important prerequisite for obtaining the jet function and hence, information about jet 
fragmentation.

	{\bf \em Decomposition of the correlation function:} Following a two source model ansatz, $C\left( {\Delta \phi } \right)$
can be written as; 
\[
C\left( {\Delta \phi } \right) = a_0\left[ H \left( \Delta \phi \right) + J\left( \Delta \phi \right) \right],
\]
where $H\left( \Delta \phi \right)$ is a second harmonic function having an amplitude  
$p_2 = v_2(\mathrm{assoc}) \times v_2(\mathrm{trig})$ and $J\left( \Delta \phi \right)$ 
is the jet function \cite{Ajitanand_05}. Here, $v_2(\mathrm{assoc})$ and 
$v_2(\mathrm{trig})$ are the amplitudes of the harmonic distributions of trigger and 
associated hadrons (respectively) relative to the azimuth of the reaction plane $\Psi_{RP}$, 
ie. $v_2(\mathrm{assoc,trig}) = \left\langle cos \left(2.(\phi_{\mathrm{assoc,trig}}-\Psi_{RP}) \right) \right\rangle$.
For these measurements, $\Psi_{RP}$ was obtained from Beam-Beam counters 
having $\eta \pm 3.5$. To obtain $J\left( \Delta \phi \right)$ one 
simply uses the value $a_0$, obtained by assuming that the magnitude of the 
Jet function $J\left( \Delta \phi \right)$ is zero at its minimum 
(ZYAM assumption) \cite{Ajitanand_05}, and subtracts 
the measured values of $H\left( \Delta \phi \right)$.

{\bf \em Extinction Method: } One can also obtain $J\left( \Delta \phi \right)$ via extinction of the 
the harmonic function $H\left( \Delta \phi \right)$ \cite{Ajitanand_05}. This is done by aligning the trigger 
hadron within an appropriately chosen angular range $\Delta \phi _c$, perpendicular to the reaction 
plane. The harmonic amplitude for the trigger particle in this configuration $v_2^{out}(\mathrm{trig})$, 
is given as \cite{Bielcikova_04}
\[
v_2^{out}(\mathrm{trig}) =\left( {\frac{2v_2 \left( {\Delta \phi _c } \right)-\sin \left( 
{2\Delta \phi _c } \right)\left\langle {\cos \left( {2\Delta \Psi _R } 
\right)} \right\rangle +\frac{v_2 }{2}\sin \left( {4\Delta \phi _c } 
\right)\left\langle {\cos \left( {4\Delta \Psi _R } \right)} \right\rangle 
}{2\left( {\Delta \phi _c } \right)-2v_2 \sin \left( {2\Delta \phi _c } 
\right)\left\langle {\cos \left( {2\Delta \Psi _R } \right)} \right\rangle 
}} \right).
\]
Thus, it is clear that $v_2^{out}(\mathrm{trig})$ can be made to vanish 
($v_2^{out}(\mathrm{trig}) \sim 0)$ by appropriate choice of $\Delta \phi _c$ for 
a particular reaction plane resolution $\Delta \Psi _R$.

\begin{figure}
\begin{minipage}[t]{0.5\linewidth}
\includegraphics[width=1.\linewidth]{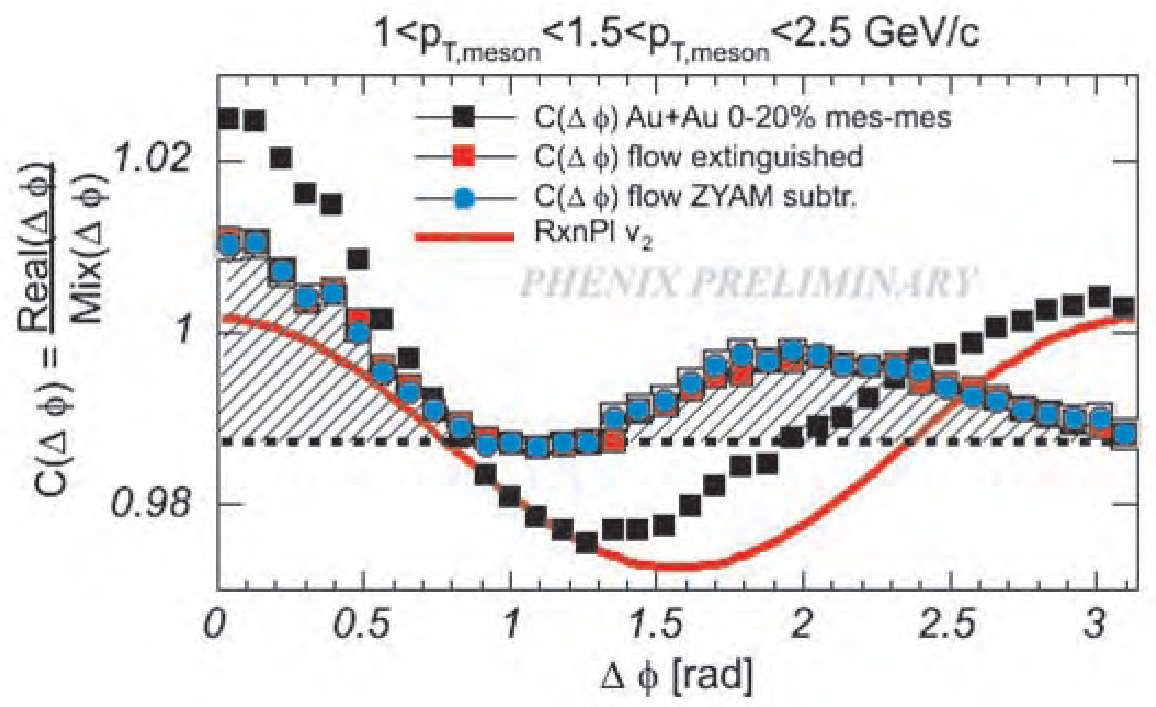}
\vskip -0.8cm
\caption{\small{ Correlation function for meson-meson pairs 
with $1.< p_T( \mathrm{assoc. meson}) 1.5 < p_T( \mathrm{trig. meson}) < 2.5$~GeV/c (black squares). 
The red curve shows the harmonic contribution. The blue circles and red squares show 
the jet correlations which result from subtraction and harmonic extinction respectively (see text).}}
\label{jet_pair_mm}
\end{minipage}
\hskip 0.2cm
\begin{minipage}[t]{0.5\linewidth}
\includegraphics[width=1\linewidth]{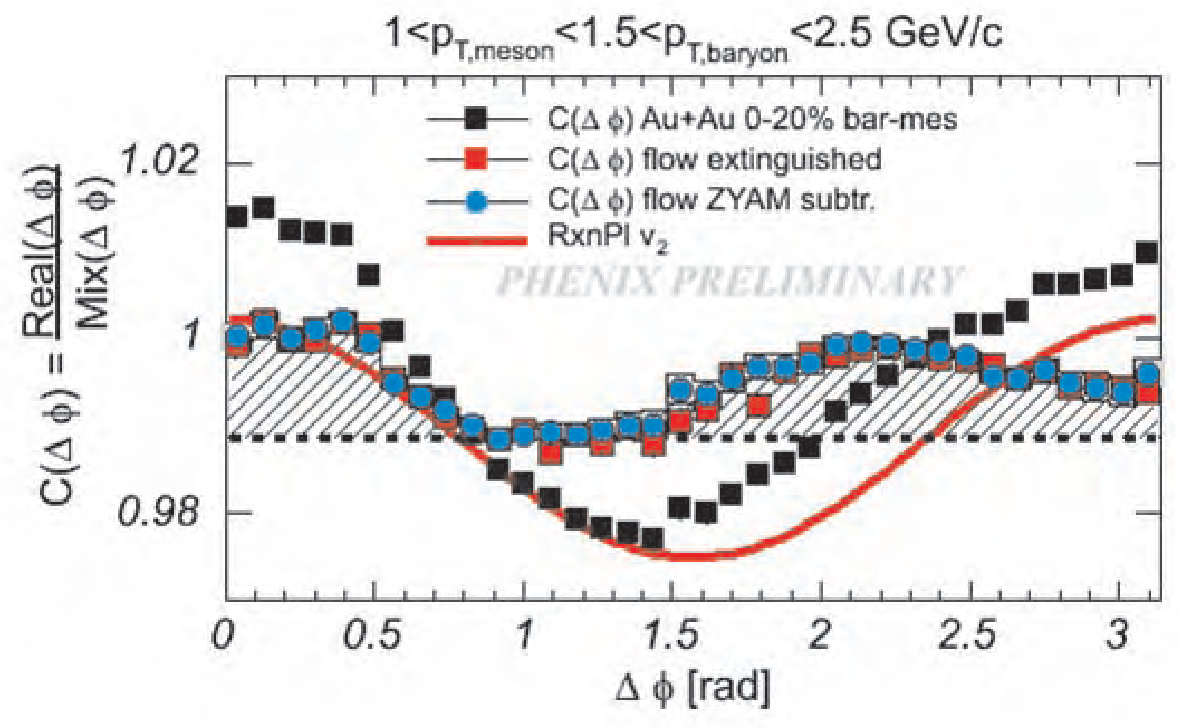}
\vskip -0.8cm
\caption{\small{Same as Fig.~\ref{jet_pair_mm}, but for baryon-meson pairs.}}
\label{jet_pair_bm}
\vskip 0.3cm
\end{minipage}
\end{figure}

{\bf \em Results: } Figures \ref{jet_pair_mm} and \ref{jet_pair_bm} show the application of these decomposition 
methods to meson-meson and baryon-meson correlation functions. They show that the  
jet-pair distributions, obtained after subtraction or extinction of the harmonic contributions, 
are significantly broadened on the away-side, independent of whether or not the trigger hadron is 
a baryon or meson. We attribute such broadening to the strong interactions between scattered 
partons and the high energy density medium.

\begin{figure}
\begin{minipage}[t]{0.5\linewidth}
\includegraphics[width=1\linewidth]{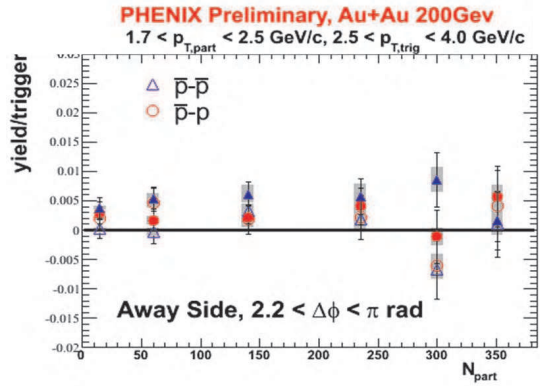}
\vskip -0.8cm
\caption{\small{Per trigger yields vs. centrality for near-side jets
obtained from p-p and p-$\overline{p}$ pairs.}}
\label{p_pbar_near}
\hskip 0.2cm
\end{minipage}
\begin{minipage}[t]{0.5\linewidth}
\includegraphics[width=1.\linewidth]{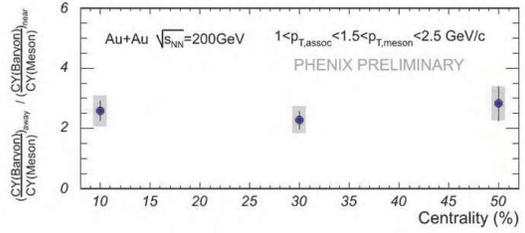}
\vskip -0.6cm
\caption{\small{ Centrality dependence of the double ratio $DBR$ }}
\label{yield_ratio}
\vskip 0.3cm
\end{minipage}
\end{figure}

It is straightforward to evaluate the conditional or per trigger yields (CY) for the 
near- and away-side jets via the integrated pair fractions indicated by the hatched area in 
Figs.~\ref{jet_pair_mm} and \ref{jet_pair_bm} \cite{Ajitanand_05} or by building per trigger yield
distributions. Fig.~\ref{p_pbar_near} compares proton and anti-proton jet yields obtained for 
near-side jets via the latter technique. The near-side yield is non-zero only for baryon-anti-baryon 
pairs, suggesting that baryon number conservation may play an important role in jet fragmentation.

One can also compare the baryonic and mesonic content of the near- and away-side jets by 
comparing the ratio CY(baryon)/CY(meson) for the near- and away-side jets. Fig.~\ref{yield_ratio} 
shows the double ratio
\[
DBR = \left( CY(baryon)/CY(meson)\right)_{away-jet}/\left( CY(baryon)/CY(meson)\right)_{near-jet}
\]
for central~-~mid-central collisions. The observed ratio of $\sim 2.5$ suggests that the away side jet is more 
baryon rich than the near side-side jet, possibly because of a modification to the fragmentation function 
of the away-side jet.

{\bf \em Three particle correlations: } Three-particle correlation functions consisting 
of a trigger hadron from the range $2.5< p_T <$4.0 GeV/c (hadron \#1) and two associated hadrons from 
the range $1.0< p_T <$2.5 GeV/c (hadron \#2 and \#3) were also studied. Correlation surfaces were constructed by way of $\Delta\phi_{1,2}$ and $\Delta\phi_{1,3}$ distributions.
For these correlation functions, the harmonic extinction method was used and 
the mixed background was created by combining the high $p_T$ particle from the real events
with associated particles from two different events.
 The correlation functions therefore contain both triples and doubles contributions. 

\begin{figure}
\begin{minipage}[t]{0.31\linewidth}
\includegraphics[width=1.\linewidth]{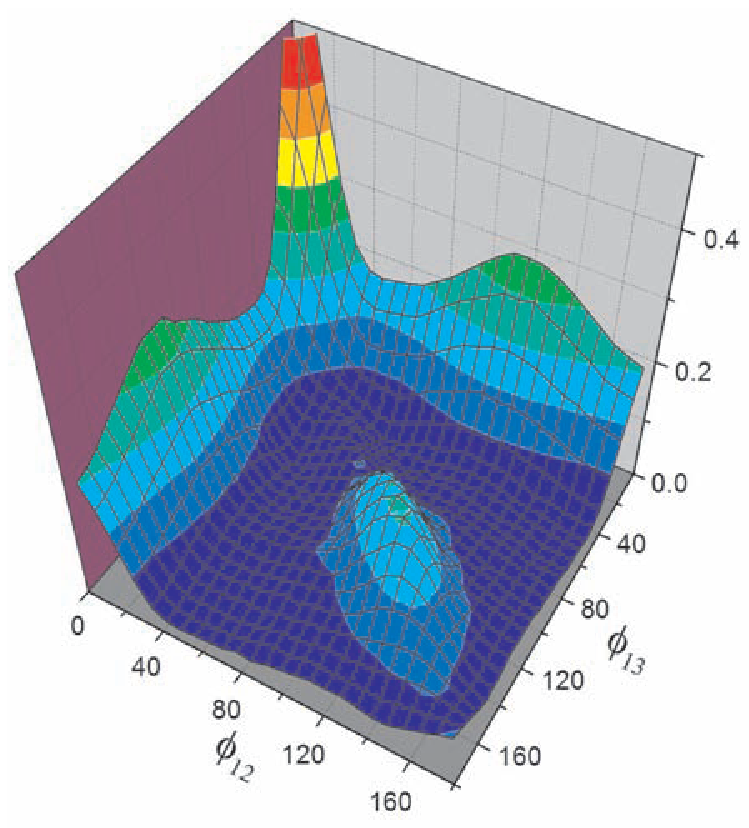}
\vskip -1.0cm
\caption{\small{ Simulated 3-particle correlations for ``bent" jets.}}
\label{3pc_sim_bjet}
\end{minipage}
\hskip 0.2cm
\begin{minipage}[t]{0.31\linewidth}
\includegraphics[width=1.\linewidth]{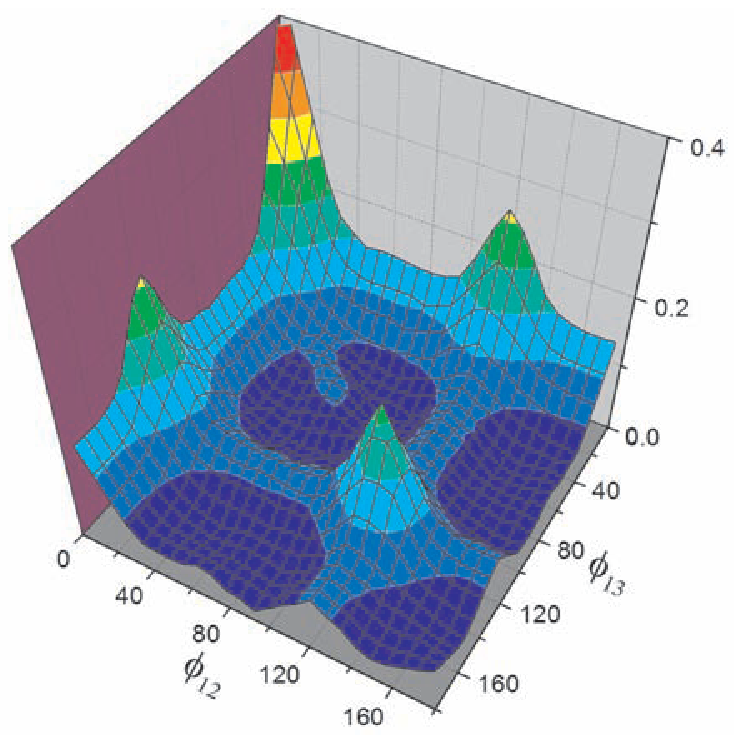}
\vskip -1.0cm
\caption{\small{ Simulated 3-particle correlations for Cherenkov jets or conical flow.}}
\label{3pc_sim_cjet}
\end{minipage}
\hskip 0.2cm
\begin{minipage}[t]{0.31\linewidth}
\includegraphics[width=1.\linewidth]{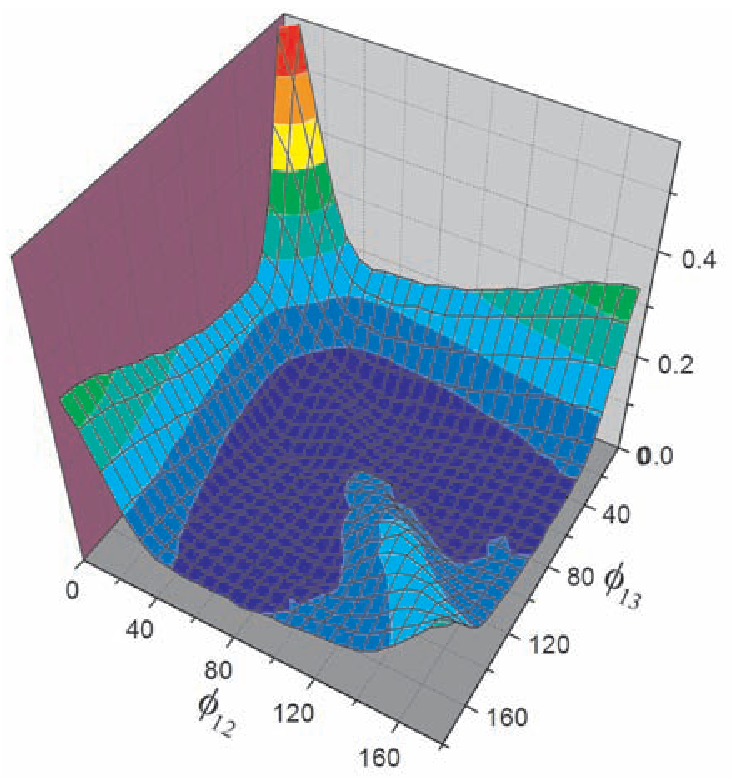}
\vskip -1.0cm
\caption{\small{ Simulated 3-particle correlations for ``normal" jets.}}
\label{3pc_sim_njet}
\end{minipage}
\hskip 0.2cm
\end{figure}

Simulated three-particle correlation surfaces 
($\Delta\phi_{1,2}$ vs. $\Delta\phi_{1,3}$) are shown in 
Figs.~\ref{3pc_sim_bjet}~-~\ref{3pc_sim_njet} for three distinct away-side jet scenarios; 
(i) a ``normal jet" in which the away-side jet axis is aligned with the leading jet axis with a spread, 
(ii) a ``bent jet" in which the away-side jet axis is misaligned by $60^o$, and 
(iii) a ``Cherenkov or conical jet" in which the leading and away-side jet axes are aligned
but fragmentation is confined to a very thin hollow cone with a half angle of $60^o$. 
The simulated results show relatively clear distinguishing features for the three scenarios
considered.

\begin{figure}
\begin{minipage}[t]{0.31\linewidth}
\includegraphics[width=1.\linewidth]{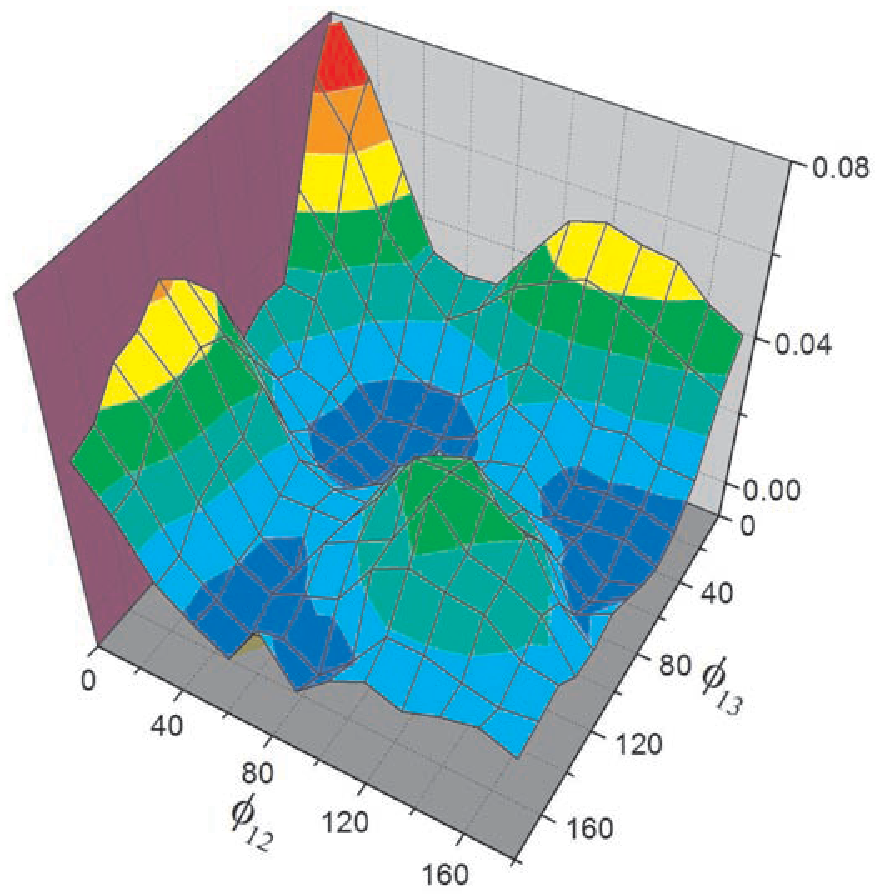}
\vskip -0.8cm
\caption{\small{ Hadron-hadron-hadron correlation function.}}
\label{3pc_data_hhh}
\end{minipage}
\hskip 0.2cm
\begin{minipage}[t]{0.31\linewidth}
\includegraphics[width=1.\linewidth]{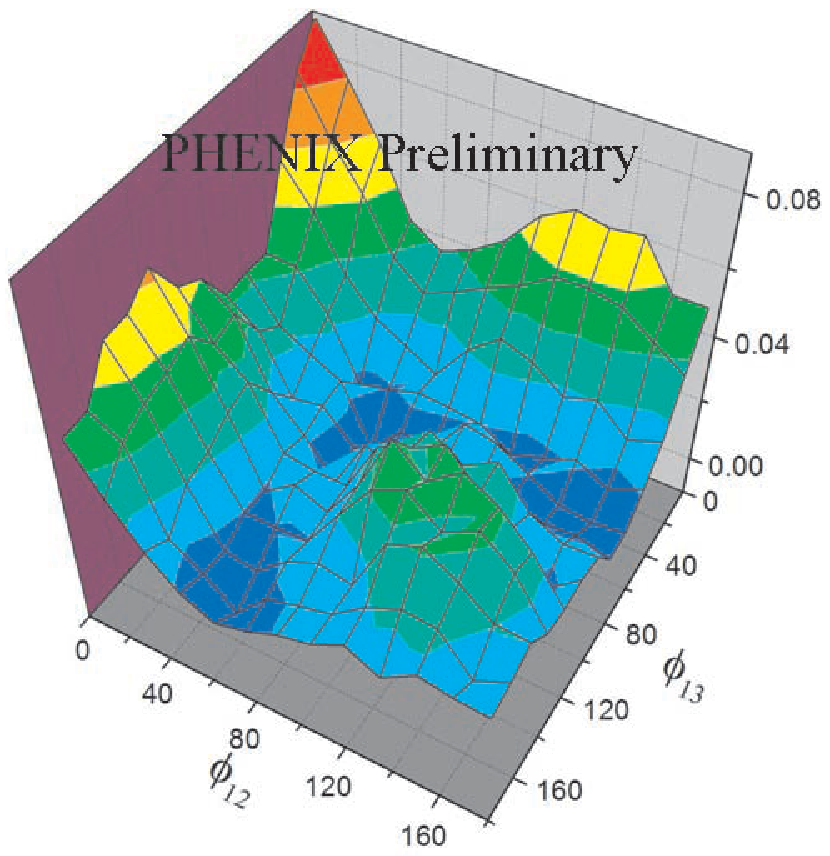}
\vskip -0.8cm
\caption{\small{ Hadron-meson-meson correlation function.}}
\label{3pc_data_hmm}
\end{minipage}
\hskip 0.2cm
\begin{minipage}[t]{0.31\linewidth}
\includegraphics[width=1.\linewidth]{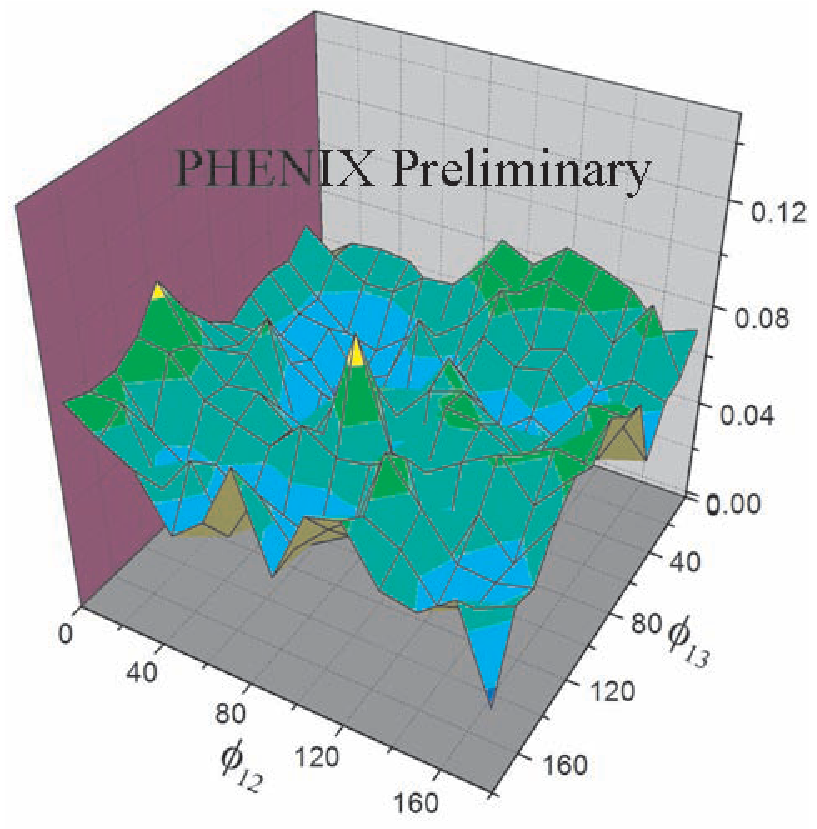}
\vskip -0.8cm
\caption{\small{ Hadron-baryon-baryon correlation function.}}
\label{3pc_data_hbb}
\end{minipage}
\hskip 0.2cm
\begin{minipage}[t]{0.31\linewidth}
\end{minipage}
%
\vskip 0.3cm
%
Further detailed quantitative investigations are however required to firm up 
the signatures in the data which distinguish between a ``bent jet" and 
a ``conical jet".

{\bf \em Summary:} Novel methodologies have been developed to remove harmonic 
contributions and extract jet functions from azimuthal correlation functions. 
Jet function and yields show strong dependence on particle flavor. The near-side jet yield is 
non-zero only for baryon-anti-baryon pairs. The jet landscape for three particle 
correlations have been obtained as a function of particle flavor. They provide 
further compelling evidence for strong modification of the away-side jet.

\end{figure}

The correlation surfaces obtained from data for the centrality selection 10-20 {\%} are shown 
in Figs.~\ref{3pc_data_hhh}~-~\ref{3pc_data_hbb}. They show a strong dependence on the flavor (PID) 
of the associated particle and clearly do not follow the expected patterns for a ``normal jet". 
We conclude that these three-particle correlation surfaces provide additional compelling evidence 
for strong modification of the away-side jet.
%
%
%
\end{document}